\documentclass[aps,prb,twocolumn,floatfix,superscriptaddress]{revtex4}

\usepackage{amsmath}
\usepackage{amssymb}
\usepackage{times}
\usepackage{braket}
\usepackage[pdftex]{graphicx}
\usepackage{blindtext}
\usepackage{hyperref}
\usepackage{cleveref}

\renewcommand{\vec}[1]{\boldsymbol{#1}}

\usepackage{xcolor}
\newcommand{\bcor}{\color{black} }

\newcommand{\ecor}{\color{black}}

\begin{document}

\title{Electrical writing, deleting, reading, and moving of magnetic skyrmioniums in a racetrack device}

\author{B{\"o}rge G{\"o}bel}
\thanks{These authors contributed equally}
\email[Corresponding author. ]{bgoebel@mpi-halle.mpg.de}
\affiliation{Max-Planck-Institut f\"ur Mikrostrukturphysik, D-06120 Halle (Saale), Germany}

\author{Alexander F. Sch{\"a}ffer}
\thanks{These authors contributed equally}
\email[Corresponding author. ]{alexander.schaeffer@physik.uni-halle.de}
\affiliation{Institut f\"ur Physik, Martin-Luther-Universit\"at Halle-Wittenberg, D-06099 Halle (Saale), Germany}

\author{Jamal Berakdar}
\affiliation{Institut f\"ur Physik, Martin-Luther-Universit\"at Halle-Wittenberg, D-06099 Halle (Saale), Germany}

\author{Ingrid Mertig}
\affiliation{Max-Planck-Institut f\"ur Mikrostrukturphysik, D-06120 Halle (Saale), Germany}
\affiliation{Institut f\"ur Physik, Martin-Luther-Universit\"at Halle-Wittenberg, D-06099 Halle (Saale), Germany}

\author{Stuart S. P. Parkin}
\affiliation{Max-Planck-Institut f\"ur Mikrostrukturphysik, D-06120 Halle (Saale), Germany}

\date{\today}

\begin{abstract}
A magnetic skyrmionium (also called 2$\pi$-skyrmion) can be understood as a skyrmion --- a topologically non-trivial magnetic whirl --- which is situated in the center of a second skyrmion with reversed magnetization. Here, we propose a new optoelectrical writing and deleting mechanism for skyrmioniums in thin films, as well as a reading mechanism based on the topological Hall voltage. Furthermore, we point out advantages for utilizing skyrmioniums as carriers of information in comparison to skyrmions with respect to the current-driven motion. We simulate all four constituents of an operating skyrmionium-based racetrack storage device: creation, motion, detection and deletion of bits. The existence of a skyrmionium is thereby interpreted as a `1' and its absence as a `0' bit.
\end{abstract}


\maketitle
Magnetic skyrmions~\cite{bogdanov1989thermodynamically,muhlbauer2009skyrmion, nagaosa2013topological} are whirl-like quasiparticles that are under consideration as carriers of information in modern data storages: Sampaio \textit{et al.}~\cite{sampaio2013nucleation} proposed to write and move skyrmions in thin film nanowires what constitutes a derivative of a racetrack storage device, initially proposed for domain walls in a ferromagnetic thin films~\cite{parkin2004shiftable,parkin2008magnetic,parkin2015memory}. The low driving current, small size and high stability of skyrmions, combined with the stackability of these tracks into three dimensions may lead to the development of highly efficient magnetic memory-storage devices with capacities that rival those of magnetic hard-disk drives, satisfying the ever-growing demand for data storage. 

Since the initial discovery of skyrmions in the form of periodic lattices in bulk single crystals of MnSi~\cite{muhlbauer2009skyrmion}, scientific effort has led to promising advances towards the utilization of isolated skyrmions as information carriers~\bcor\cite{sampaio2013nucleation,romming2013writing, maccariello2018electrical,hamamoto2016purely,jiang2017direct, kang2016skyrmion,kang2016voltage}\ecor. Still, one major issue of driving skyrmions on a racetrack is the skyrmion Hall effect~\cite{nagaosa2013topological,zang2011dynamics,jiang2017direct,litzius2017skyrmion} originating from the real-space topological properties of skyrmions. A skyrmion carries a topological charge of $N_\mathrm{Sk}\pm 1$, defined as the integral over the topological charge density
\begin{align}
n_\mathrm{Sk}(\vec{r})  = \frac{1}{4\pi} \vec{m}(\vec{r}) \cdot \left[ \frac{\partial \vec{m}(\vec{r})}{\partial x}  \times  \frac{\partial \vec{m}(\vec{r})}{\partial y}  \right]\label{eq:nsk},
\end{align}
where $\vec{m}(\vec{r})$ is the unit vector magnetization field. Skyrmions driven by spin-polarized electrical currents are not propelled parallel to the racetrack. They experience a transverse deflection to the edge of the racetrack where they may be confined. This effect is detrimental for racetrack applications.
 
Theoretical suggestions for suppressing the skyrmion Hall effect are to manipulate the driving torque orientation~\cite{zhang2015skyrmion,kim2018asymmetric,buttner2018theory, gobel2018overcoming} or to use antiferromagnetic skyrmions with a vanishing topological charge~\cite{barker2016static,zhang2016magnetic,zhang2016antiferromagnetic,gobel2017afmskx} instead of skyrmions. However, both approaches have not yet been realized experimentally.

\begin{figure*}
	\centering
	\includegraphics[width=1\textwidth]{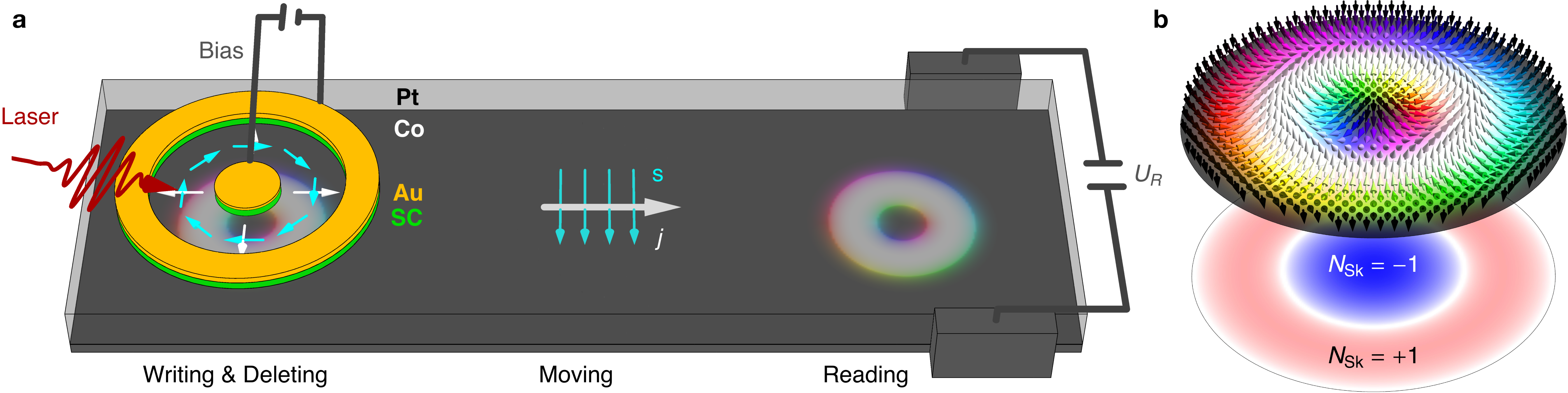}                
	\caption{\textbf{Skyrmionium-based racetrack storage device.} \textbf{a}, Schematic presentation of the proposed device including the four constituents: writing, deleting, moving and reading. A skyrmionium, the circular object in the Co layer (gray), is written or deleted by a photosensitive switch built from gold (Au) and a semi-conductor (SC). A radial current (white) is triggered upon illuminating the antenna with a fs-laser pulse due to the applied bias voltage. The current mainly flows in the Pt layer (transparent) where the SHE injects spins (cyan) into the Co layer that are oriented perpendicularly to the plane normal and current directions. Depending on the polarity of the gate voltage, the sign of the optically activated current pulse and the orientation of the polarization are determined, so that a skyrmionium can be written or deleted. To move the skyrmionium a uniform current density $\vec{j}$ is applied along the track, again generating spins $\vec{s}$ that exert a SOT onto the skyrmionium. When a skyrmionium is located near the two leads on the right, a Hall voltage $U_R$ can be measured, allowing for a distinct detection of a skyrmionium bit. \textbf{b}, (top) Magnetic texture of a skyrmionium, and (bottom) topological charge density $n_\mathrm{Sk}$ with opposite signs for the inner skyrmion and the outer ring. }
	\label{fig:skyrmionium}
\end{figure*}

Here, we utilize another type of magnetic quasi-particle (Fig.~\ref{fig:skyrmionium}a) with a zero topological charge: the skyrmionium (also called a $2\pi$-skyrmion)~\bcor\cite{bogdanov1999stability,beg2015ground,zhang2016control, finazzi2013laser,zheng2017direct,zhang2018real,hagemeister2018controlled, kolesnikov2018skyrmionium,li2018dynamics,shen2018motion, pylypovskyi2018chiral}\ecor. The skyrmionium has been observed experimentally created by laser pulses~\cite{finazzi2013laser}, as target skyrmionium in nanodiscs~\cite{zheng2017direct} and very recently in a thin ferromagnetic film on top of a topological insulator~\cite{zhang2018real}. A magnetic skyrmionium (Fig.~\ref{fig:skyrmionium}b) can be described as a skyrmion, with a second skyrmion situated in the center. The inner skyrmion has a reversed polarity and deforms the outer skyrmion to a ring. 

Here, we show that skyrmioniums can be used as carriers of information in a racetrack storage device (Fig.~\ref{fig:skyrmionium}a): the existence of a skyrmionium is interpreted as a `1' bit, while its absence is a `0' bit.
Based on recent progresses in optically generated current pulses~\cite{yang2017ultrafast} we propose a way to write and delete magnetic skyrmioniums on the ps timescale, so that the current-induced skyrmionium flow --- without the detrimental skyrmion Hall effect --- can remain steady while writing. Also we show that skyrmioniums can be detected electrically by their topological Hall signal, that arises from the local topological charge density (Fig.~\ref{fig:skyrmionium}b bottom), \bcor even though the global topological charge vanishes.\ecor\\
\\
\textbf{Results}\\
\textbf{Skyrmionium racetrack.}
In the following, we simulate and analyze point by point the four essential constituents to operate a racetrack-storage device, based on magnetic skyrmioniums. First, we show via micromagnetic simulations how skyrmioniums can be written and deleted by optically excited localized current pulses. 
Thereafter, we present advantages in the current-driven motion of skyrmioniums compared to that of conventional skyrmions; we explain the simulated results by an effective description using the Thiele equation. Ultimately, we show via Landauer-B\"uttiker calculations that the local separation of the two subskyrmions of a skyrmionium can be exploited to electrically detect skyrmioniums even though they exhibit no topological Hall voltage when integrated over the whole sample.

For the device (Fig.~\ref{fig:skyrmionium}a) we consider a magnetic layer on a heavy metal: here we exemplarily select  Co (gray) on Pt (transparent), as in Refs.~\onlinecite{sampaio2013nucleation,zhang2016control}. In this setup an applied charge current density $\vec{j}$ (white) within the Pt layer is translated  to a spin current by the considerable spin Hall effect that has been observed in Pt. The spin current flows perpendicular to the plane into the Co potentially hosting skyrmioniums. The spin polarization $\vec{s}$ (cyan) is perpendicular to $\vec{j}$ and the plane normal, thereby leading to a spin-orbit torque~\cite{slonczewski1996current} (SOT) that can propel a skyrmionium.

On top of the basic racetrack that is formed from the Pt/Co bilayer a photosensitive switch~\cite{ketchen1986generation} is fabricated that can switch the magnetization as experimentally shown in Ref.~\onlinecite{yang2017ultrafast}. A circular gold disk inside a gold ring is isolated from the metallic racetrack by an underlying semiconducting layer (green) so that a bias voltage can be applied between the inner and outer gold electrodes (Corbino geometry). The semiconductor is electrically activated by fs-laser pulses which generate a radially symmetric current pulse profile \bcor$\vec{j}=j_\mathrm{write} \frac{r_0}{r}\,\hat{e}_r$ \ecor (white) in the Pt-layer, that, by analogy with the explanation given above, leads to a toroidal spin polarization profile \bcor$\vec{s}(\vec{r})\parallel \pm\hat{e}_\varphi$ \ecor (cyan) of the spin currents \bcor(parallel to $\hat{e}_z$) \ecor that diffuse into the Co layer to create or delete the skyrmionium. To model the skyrmionium generation, deletion and motion we use a micromagnetic framework based on the Landau-Lifshitz-Gilbert (LLG) equation~\cite{landau1935theory,gilbert1955lagrangian,slonczewski1996current}. For details and simulation parameters see Methods.

For the reading process we utilize the local topological properties of a skyrmionium. The non-zero topological charge density $n_\mathrm{Sk}$ leads to a deflection of electrons in a transverse direction. A Hall voltage $U_R$ builds up that can be measured by attaching two small leads (gray) to the sides of the device. \\
\\
\textbf{Optoelectrical writing of skyrmioniums.}
Several mechanisms for writing skyrmions have been proposed, such as the application of spin-polarized currents~\cite{fert2013skyrmions}, laser beams~\cite{finazzi2013laser} and electron beams~\cite{schaffer2017ultrafast,schaffer2017ultrafast2}. These mechanisms can potentially be adapted to generate also skyrmioniums. \bcor It has been predicted that skyrmioniums can be generated by alternating the out-of-plane orientation of an external magnetic field \cite{hagemeister2018controlled} or by the perpedicular injection of spin currents  \cite{zhang2016control,kolesnikov2018skyrmionium}. \ecor
\begin{figure*}
  \centering
  \includegraphics[width=2\columnwidth]{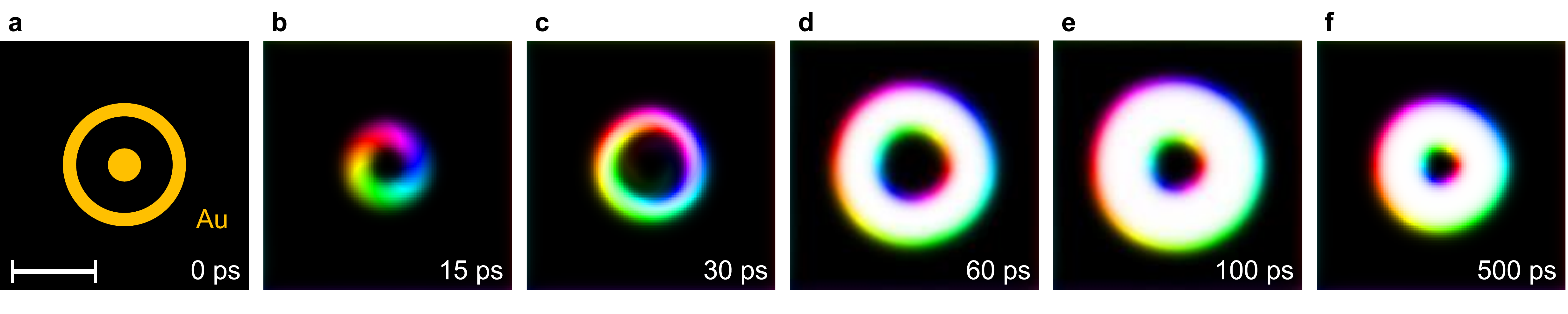}              
  \caption{\textbf{Writing skyrmioniums by optically excited current pulses}. \textbf{a}, The starting point is a ferromagnet magnetized along $-z$. The orange areas indicate the gold nanostructures that generate a current pulse of  $9\,\mathrm{ps}$ duration (FWHM) with $j_\mathrm{max}(t=15\,\mathrm{ps})=2\times 10^{13}\,\mathrm{A}/\mathrm{m}^2$ at the disk. \textbf{b}, After the spin current has induced a skyrmionium-shaped excitation the quasiparticle relaxes in \textbf{c-f}. The color code is the same as in Fig.~\ref{fig:skyrmionium}. An animated version is accessible in Supplementary Video 1. For sample and beam parameters see text and Methods. {\bcor The scale bar corresponds to $50\,\mathrm{nm}$.}}
  \label{fig:write}
\end{figure*}
\begin{figure*}
  \centering
  \includegraphics[width=2\columnwidth]{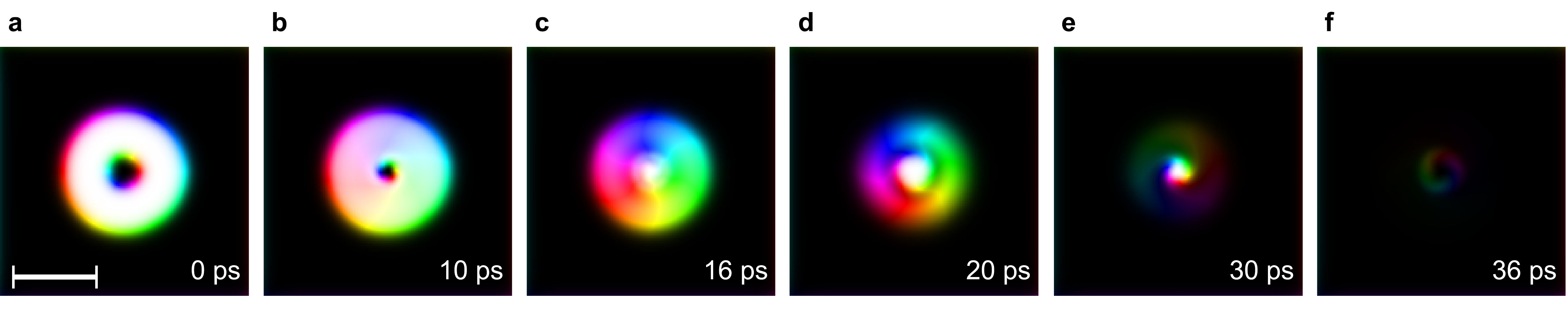}  
  \caption{\textbf{Deleting skyrmioniums by optically excited current pulses}. \textbf{a}, We start from the skyrmionium stabilized in Fig.~\ref{fig:write} and generate a spin current with opposite polarization ($\hat{e}_\varphi$ direction) compared to the writing process achieved by switching the sign of the bias voltage. \textbf{b,c}, The domain wall unwinds the inner skyrmion leading to its collapse. \textbf{d,e,f}, Subsequently, the skyrmion-like configuration contracts and dissolves. An animated version is accessible in Supplementary Video 2. {\bcor The scale bar corresponds to $50\,\mathrm{nm}$.}}
  \label{fig:delete}
\end{figure*}

\bcor We propose a faster writing mechanism, where spins are injected from the perpendicular direction for an ultrashort duration. We consider \ecor a SOT-driven approach based on a nanostructured skyrmionium manipulation unit as sketched in Fig.~\ref{fig:skyrmionium}a. Optically excited radially symmetric charge currents (white) in the Pt-layer lead to out-of-plane spin currents with a controllable toroidal spin polarization configuration (cyan): 
For opposite signs of the applied bias-voltage opposite spin polarizations are achieved ($\pm\hat{e}_\varphi$). According to Yang \textit{et al.}~\cite{yang2017ultrafast} current densities on the order of $j_\mathrm{max}=2\times 10^{13}\,\mathrm{A}/\mathrm{m}^2$ could be created for a pulse duration of $9\,\mathrm{ps}$ full width at half maximum (FWHM). These values \bcor were adapted in our proposed optoelectrical writing and deleting process of a single skyrmionium. They cannot be reached by a conventional perpendicular spin current injection. \ecor The diameter of the inner disk of the photosensitive switch is $20\,\mathrm{nm}$ and the outer ring's inner diameter is $60\,\mathrm{nm}$ in order to match the skyrmionium's dimension. These dimensions are at the limit of what is possible today using conventional lithographic processes.

Starting from an initially uniform magnetization pointing into the $-z$ direction (Fig.~\ref{fig:write}a), the system is excited by an out-of-plane spin current with toroidal polarization. Because of the ultrashort current pulses of $9\,\mathrm{ps}$ (FWHM) the excitation itself is non-adiabatic and the magnetic texture will relax on a longer time scale.
 
During the current-pulse (Fig.~\ref{fig:write}b, maximum current at $15~$ps) the magnetization in the excited ring-shaped region begins to follow the spin current's polarization, i.e. in the $-\hat{e}_\varphi$ direction. The amplitude of the current as well as its location needs to be tuned in such a way, that it will effectively switch a ring-shaped domain of a suitable size (Fig.~\ref{fig:write}c). 
Subsequently not only spin waves propagate radially, but also the central circular region remaining in its initial orientation starts to pulsate (Fig.~\ref{fig:write}c,d,e,f). Associated with this, the domain wall between the central $-z$ region and the intermediate $+z$ region is rotating such that a central N\'eel-skyrmion is generated, thereby, in total, constituting a skyrmionium. The slowest relaxation is the adjustment of the skyrmionium's size, taking place after the central fluctuations decay. The shrinking towards the final diameter (around $80\,$nm) lasts for $\sim 500\,$ps (Fig.~\ref{fig:write}e,f). \\
\\
\textbf{Optoelectrical deleting of skyrmioniums.}
Deleting a non-collinear magnetic texture means turning it into a ferromagnetic state. Since no stabilizing external magnetic field is applied to the racetrack, the magnetization can in principle point into both out-of-plane directions. The uncontrolled annihilation of a skyrmionium can therefore easily lead to a local reversal of the magnetization direction, i.\,e., the formation of a domain. Therefore `1' bits need to be turned into `0' bits in a controlled way; no ferromagnetic domains must form. An efficient way is to invert the writing mechanism by reversing the bias voltage, which goes along with a change of the spin current's polarization from $-\hat{e}_\varphi$ to $+\hat{e}_\varphi$. 

The annihilation process is shown in Fig.~\ref{fig:delete}. The generated spin current effectively unwinds the skyrmionium structure step by step. First, the rotation of the domain wall leads to the dissolving of the central $-z$ domain (Fig.~\ref{fig:delete}a,b,c). Second, the remaining skyrmion-like configuration contracts (Fig.~\ref{fig:delete}d,e) until it collapses (Fig.~\ref{fig:delete}f). The system relaxes towards the ferromagnetic state in less than $40~$ps.\\
\\
\textbf{Current-driven motion of skyrmioniums.}
Skyrmions and skyrmioniums can be driven by spin torques. As discussed above, we use a two-layer setup that utilizes SOT, which means $\vec{s}\parallel$\bcor$-$\ecor$\vec{y}$ for $\vec{j}\parallel\vec{x}$. This mechanism has been proven to be far more efficient compared to propagation induced by spin-polarized currents applied within the ferromagnetic layer (spin-transfer torque)~\cite{sampaio2013nucleation,zhang2016control}.
\begin{figure*}
  \centering
	\includegraphics[width=\textwidth]{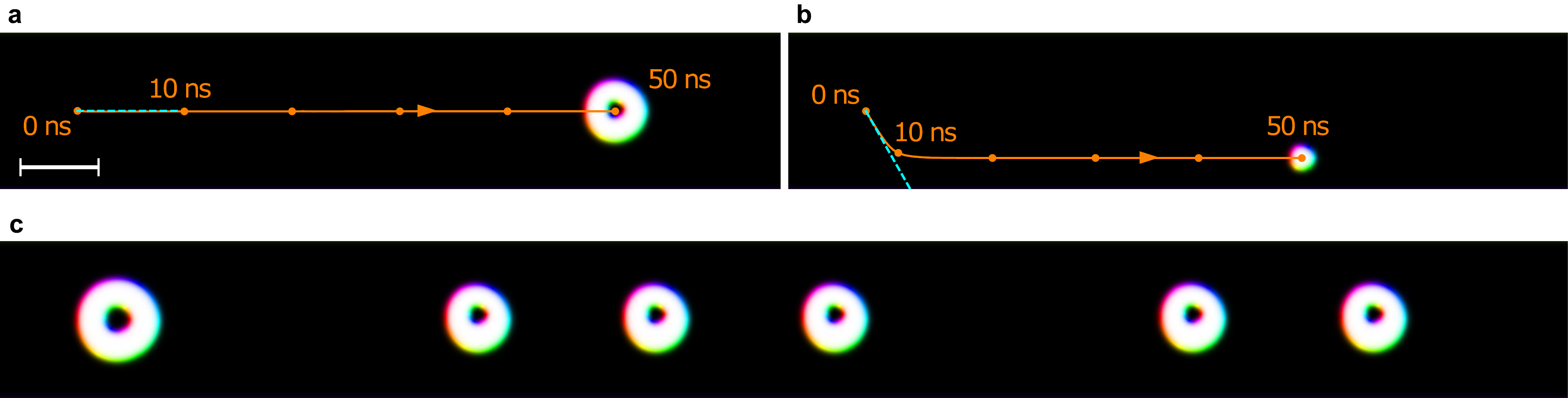}    
  \caption{\textbf{Current-driven motion of skyrmioniums.} \textbf{a}, A skyrmionium in a racetrack is driven by SOT: applied current density $j_x \Theta_\mathrm{SH}=0.6\,\mathrm{MA}/\mathrm{cm} ^2$, injected spins are oriented along \bcor $-$\ecor$y$. \textbf{b}, The motion of a skyrmion is shown for comparison. In both cases the image is taken after $50\,\mathrm{ns}$ of propagation time. The orange curve shows the trajectory of the quasiparticles' centers (starting point is indicated by $0\,\mathrm{ns}$). The blue line indicates the motion direction under the skyrmion Hall angle, calculated from the Thiele equation (see text). \textbf{c}, The results of a `writing-under-current' simulation are shown after $35\,\mathrm{ns}$ propagation time in a racetrack with a doubled length. Due to the stronger current (\bcor$j_x \Theta_\mathrm{SH}=2.0\,\mathrm{MA}/\mathrm{cm} ^2$\ecor) the skyrmioniums are slightly deformed. Skyrmioniums are written after $0\,\mathrm{ns}$, $5\,\mathrm{ns}$, $15\,\mathrm{ns}$, $20\,\mathrm{ns}$, $25\,\mathrm{ns}$, and $35\,\mathrm{ns}$. The last skyrmion is not yet fully relaxed. An animated version of the skyrmionium-sequence generation is accessible in the Supplementary Video 3. {\bcor The scale bar corresponds to $100\,\mathrm{nm}$.}}
  \label{fig:move_sksk}
\end{figure*}

In our simulations (Fig.~\ref{fig:move_sksk}a,b) a reference skyrmion (Fig.~\ref{fig:move_sksk}b) first moves to the edge partially along the \bcor$-$\ecor$y$ direction of the racetrack for about $10\,$ns and then moves at a steady velocity along the confining edge along the \bcor+\ecor$x$ direction. The skyrmionium (Fig.~\ref{fig:move_sksk}a) on the other hand is propelled almost instantly to the steady state velocity and moves in the middle of the racetrack along \bcor+\ecor$x$.

The results of micromagnetic simulations can most easily be understood by an effective center-of-mass description of magnetic quasiparticles (velocity $\vec{v}$): the Thiele equation (in units of force)~\cite{thiele1973steady,sampaio2013nucleation,zhang2016control,gobel2018overcoming}
\begin{align}
b\,\vec{G}\times\vec{v}-b\underline{D}\alpha\vec{v}+Bj\underline{I}\vec{s}=\nabla U(y)\label{eq:thielesot}.
\end{align}
The properties of the respective quasiparticle are condensed into the gyromagnetic coupling vector $\vec{G}=G\vec{e}_z$ with $G=-4\pi N_\mathrm{Sk}$, and the dissipative tensor $\underline{D}$ determined by $D_{ij}=\int\partial_{i}\vec{s}(\vec{r})\cdot\partial_{j}\vec{s}(\vec{r})\,\mathrm{d}^2r$. Only $D_{xx}$ and $D_{yy}$ are nonzero. The tensor $\underline{I}$ is calculated from $I_{ij}=\int[\partial_{i}\vec{s}(\vec{r})\times\vec{s}(\vec{r})]_{j}\,\mathrm{d}^2r$ and has only nonzero $xy$ and $yx$ elements for the stabilized N\'{e}el skyrmion (the type of skyrmion is determined by the Dzyaloshinskii-Moriya interaction (DMI)~\cite{dzyaloshinsky1958thermodynamic,moriya1960anisotropic} arising at the interface between the Pt and Co layers) and skyrmionium. This tensor describes the interaction of injected spins $\vec{s}$ and the magnetic texture. The constants are $b=M_s d_z/\gamma_e$ and $B=\hbar/(2e)\Theta_\mathrm{SH}$.

While neglecting the racetrack potential $U$ (minimum in the middle of the racetrack), both textures experience a skyrmion Hall angle of 
$\theta_\mathrm{Sk}=\arctan\left(G/D_{xx}\alpha\right)$, which gives an angle of \bcor$-$\ecor$60.5^\circ$ for the skyrmion and $0^\circ$ for the skyrmionium with respect to the \bcor$+$\ecor$x$ direction, in agreement with the first period of the simulation (blue dashed lines in Figs.~\ref{fig:move_sksk}a,b). 
The magnetic quasiparticles move at a velocity of
\begin{align}
v_x=\frac{B}{b}\frac{I_{xy}}{D_{xx}}\frac{1}{\alpha}j_x-\tan\theta_\mathrm{Sk}v_y\label{eq:thielv}
\end{align}
along the racetrack. If the current density $j_x$ is small enough, a skyrmion moves to the edge of the racetrack due to its topological charge, until the gradient potential of the racetrack edge compensates the transverse force. In this case the longitudinal velocity is increased, because the second term vanishes. Due to $\theta_\mathrm{Sk}=0$ a skyrmionium on the other hand moves instantly at a constant velocity, which is given by the first term of Eq.~\ref{eq:thielv}.

In agreement with Ref.~\onlinecite{zhang2016control} we find a slightly increased skyrmionium velocity ($|v_x^\mathrm{st}|=13.8\,\mathrm{m}/\mathrm{s}$) compared to the skyrmion velocity ($|v_x^\mathrm{st}|=13.2\,\mathrm{m}/\mathrm{s}$) even in the steady state, which is explained by $v_x\propto I_{xy}/D_{xx}$ in Eq.~\ref{eq:thielv}. In infinitely wide racetracks this ratio is equal for skyrmions and skyrmioniums. In finite tracks however, the confining potential  deforms the magnetic quasiparticles slightly, altering the above ratio. Since skyrmioniums are larger than skyrmions, they experience a stronger deformation which manifests itself in a slightly increased $I_{xy}/D_{xx}$ ratio.

The striking advantage of skyrmioniums as carriers of information compared to skyrmions becomes apparent in the first $10\,\mathrm{ns}$ of their motion after a current pulse is applied. During this period of time the skyrmionium already moves at maximum speed in the middle of the track, thereby allowing the writing of several skyrmioniums in sequence while the driving current is still applied (Fig.~\ref{fig:move_sksk}c). 

{\bcor Similarly to the skyrmion-skyrmion interaction~\cite{schaeffer2019stochastic}, also the interaction between skyrmioniums is decreasing exponentially with the distance between them (see Supplementary Figure~S1). A repulsion of the quasiparticles is mainly limited to the case of a spatial overlap of the spin textures, therefore leaving the inter-skyrmionium distance in Fig.~\ref{fig:move_sksk}c constant during the considered time period.}

In Fig.~\ref{fig:move_sksk}c we apply a current density of \bcor$j\Theta_\mathrm{SH}=2.0\,\mathrm{MA}/\mathrm{cm}^2$\ecor. The skyrmionium moving at $46.28\,\mathrm{m}/\mathrm{s}$ is no longer rotationally symmetric: Its inner part is pushed to the \bcor top \ecor while the outer ring is dragged to the \bcor bottom \ecor of the racetrack, in accordance with the opposite skyrmion Hall effects that originate in the opposite partial topological charges of the two skyrmionium parts. 

The maximal current density that can be applied to skyrmions and skyrmioniums is limited to around the same value. When the driving current is too large skyrmioniums self destruct because the forces pushing the two parts of the skyrmionium in opposite directions become too large~\cite{zhang2016control} \bcor(For an analysis of the current dependence of the skyrmionium velocity and stability see Supplementary Figure~S2)\ecor. On the other hand skyrmions are annihilated at the edge. For skyrmioniums the steady state velocity can be increased up to around $140\,\mathrm{m}/\mathrm{s}$. \bcor Alternatively, skyrmioniums can also be driven by spin waves \cite{shen2018motion,li2018dynamics}.\ecor
\\
\\
\textbf{Electrical reading of skyrmioniums.}
Due to its distinct magnetization a skyrmionium can easily be detected by out-of-plane measurements. However, electrical in-plane measurement can be included in the racetrack geometry more easily. For this reason we consider detection of skyrmioniums via the Hall voltage as has been done experimentally for conventional skyrmions~\cite{maccariello2018electrical}. 
\begin{figure}
  \centering
  \includegraphics[width=1.0\columnwidth]{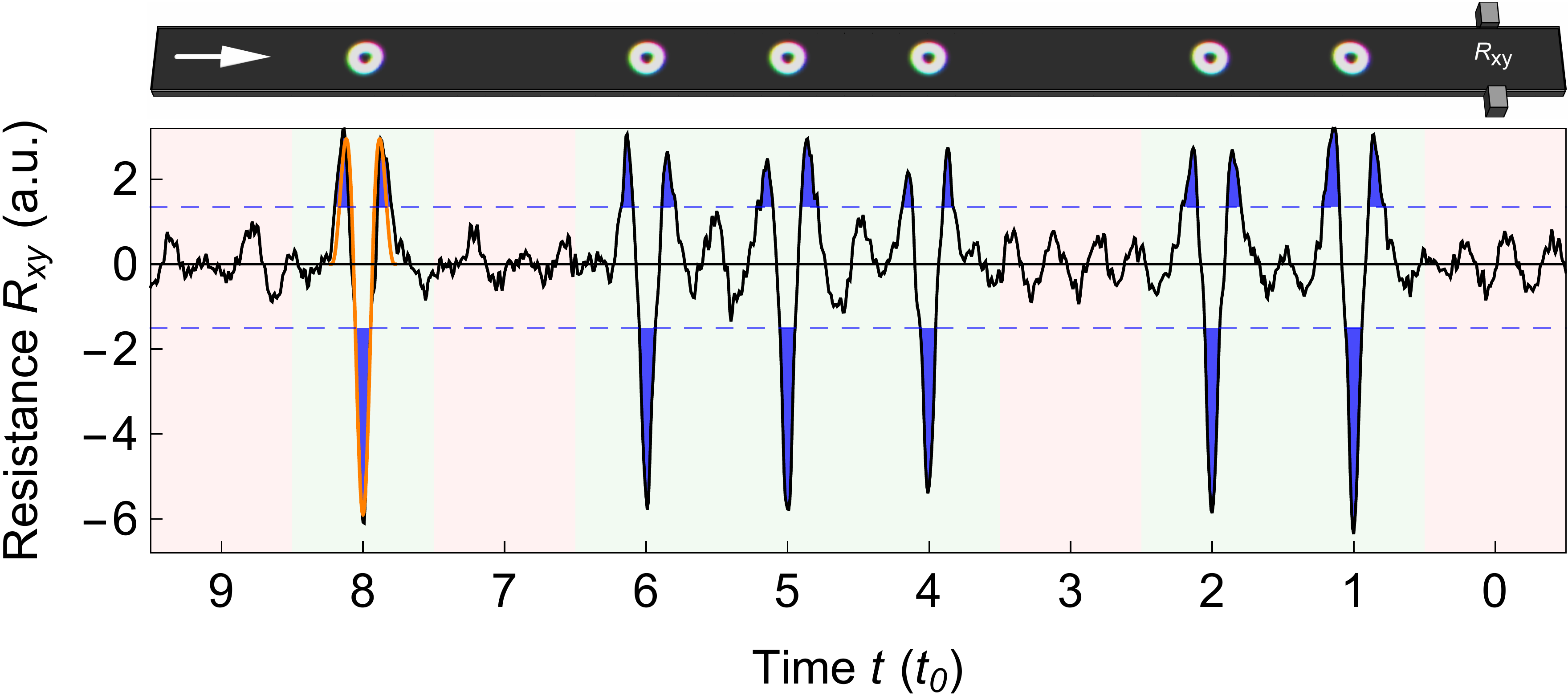}             
  \caption{\textbf{Electrical reading of skyrmioniums}. The black curve shows the calculated transverse resistance signal when the skyrmionium sequence (top panel) passes the leads. In the presence of a skyrmionium between the leads (green background) a distinct pattern is found that exceeds two thresholds (dashed lines) three times. In good approximation the signal is proportional to the integrated topological charge density between the leads (orange). If no skyrmionium is present (red) the curve fluctuates around zero. Parameters: skyrmionium radius: 40 sites, track width: 120$\times$ 2000 sites, bit width: 240 sites, lead width: 29 sites, $m/t=5$, $E_\mathrm{F}=-8.5t$.}
  \label{fig:read}
\end{figure}

When a small reading current $I$ is applied along the track the Hall resistivity is given by anomalous Hall and topological Hall contributions. The anomalous contribution is proportional to the net magnetization of the texture between the two detecting leads. For this reason every non-collinear magnetic texture is easily detected by the anomalous Hall effect. However the signal is rather similar for different textures. Skyrmioniums can not unambiguously be distinguished from skyrmions or even domain walls. This problem is resolved by the additional topological contribution to the Hall effect.

The topological Hall effect~\cite{bruno2004topological,neubauer2009topological,hamamoto2015quantized,gobel2017THEskyrmion, gobel2017QHE,hamamoto2016purely,maccariello2018electrical, nagaosa2013topological,yin2015topological} is a hallmark of the skyrmion phase: Traversing electrons are deflected into a transverse direction, since their spins (partially) align with the non-collinear texture and a Berry phase is accumulated. The topological charge density acts like a fictitious magnetic field, called an emergent field~\cite{nagaosa2013topological}. We show that even though a skyrmionium has a zero topological charge it exhibits a distinct topological Hall signal that allows for a failsafe detection of skyrmioniums as `1' bits in comparison to other non-collinear textures that may appear as defects in imperfect racetracks.

We calculate the Hall resistance for skyrmioniums in a racetrack by means of the Landauer-B{\"u}ttiker formalism~\cite{landauer1957spatial,buttiker1988absence}, by analogy with Refs.~\onlinecite{hamamoto2016purely,yin2015topological} where skyrmions have been considered (see Methods for details).
To model the interaction of electrons with the magnetic texture we considered a tight-binding model, which features nearest-neighbor hopping (amplitude $t$; creation and annihilation operator $c_i^\dagger$, $c_i$) and a Hund's coupling term (amplitude $m$, vector of Pauli matrices $\vec{\sigma}$)
\begin{align}
 H & = \sum_{\braket{i,j}} t \,c_{i}^\dagger c_{j} + m \sum_{i} \vec{m}_{i} \cdot (c_{i}^\dagger \vec{\sigma}c_{i}).\label{eq:ham}
\end{align}
Without the presence of skyrmioniums the Hamiltonian for the ferromagnet gives the energy bands $E=2t[\cos(k_xa)+\cos(k_ya)]\pm m$. Since skyrmions are detected most easily for low carrier concentrations~\cite{hamamoto2016purely} we set the Fermi energy of the system close to the lower band edge, where the electrons behave like free electrons ($E_\mathrm{F}=-8.5t$ for $m=5t$). 

Since $N_\mathrm{Sk}=0$ the topological Hall effect vanishes globally: The inner part of the skyrmion deflects electrons to the bottom, while the outer ring redirects electrons into the opposite direction. Fortunately, this spatial separation of the two opposing contributions leads to a non-zero signal in a local measurement (Fig.~\ref{fig:read}). The topological Hall resistance $R_{xy}=(U_\mathrm{up}-U_\mathrm{down})/I$ is determined by the difference in voltage $U$ at the two leads normalized by the reading current.

Whenever a skyrmionium approaches the contacts, at first only electrons deflected by the outer ring are detected. Later, when the skyrmionium is right between the leads, the inner part dominates the electron deflection and the effective charge accumulation is reversed. Finally, upon leaving the vicinity of the contacts, only the outer ring contributes to the signal. This leads to a characteristic curve that is well approximated (orange) by the topological charge density between the leads for a skyrmionium at position $x$
\begin{align}
R_{xy}(x)&\propto \int_y\int_{-x_0}^{x_0} n_\mathrm{Sk}(x-x',y')\, \mathrm{d}x'\mathrm{d}y'.
\end{align}
Electrons that traverse the spin texture are deflected by the locally nonzero emergent field of the skyrmionium $\vec{B}_\mathrm{em}\propto n_\mathrm{Sk}\vec{e}_z$ to the leads (voltage $U_\mathrm{up}$ and $U_\mathrm{down}$) of finite width ranging from $x=-x_0$ to $+x_0$. Note, that for $x_0\rightarrow\infty$ the result of zero global resistivity is recovered, independent of the position of the skyrmionium.\\
\\
\textbf{Discussion}\\
In this Paper we simulated the four fundamental constituents of a racetrack storage device utilizing magnetic skyrmioniums as carriers of information.

For the writing and deleting mechanism we proposed a new method that utilizes the optoelectrical control of localized spin currents and their polarizations. Based on previous experimental advances we designed a nanostructured geometry enabling the writing or deleting of single skyrmionium bits, depending on the sign of the bias voltage. 
Since the writing process is ultra-fast, skyrmioniums can be written while the driving current is applied even at the maximal velocity of the bits along the track of around $140\,\mathrm{m}/\mathrm{s}$ (Fig.~\ref{fig:move_sksk}c). The reliability of this deterministic method is emphasized by the result, that an excitation with the `wrong' gate voltage can not change an existing bit (see Supplementary Videos 4 and 5). In that case the spin current's associated chirality is not suited to wind or unwind the present configuration, respectively. 
Furthermore, even when room temperature fluctuations (see Methods for details) are accounted for, the proposed manipulation technique still works (see \bcor Supplementary Figure S3 \ecor and Supplementary Videos 6 and 7) \bcor what makes the presented mechanism highly attractive over other proposals. Also, we checked the range of parameters characterizing the optoelectrical writing mechanism that allow for a controlled generation of skyrmioniums (cf. Supplementary Figure~S4).\ecor

We analyzed the motion of skyrmioniums under application of electrical currents in the Pt layer where a spin current is injected perpendicularly into the Co layer (SOT). Due to their vanishing topological charge, skyrmioniums move in the middle of the racetrack and reach a steady state of motion almost instantly.

Reading magnetic skyrmioniums is possible via measurements of the Hall voltage. A local drop in the net magnetization leads to the emergence of an anomalous Hall effect and the segregation of the two skyrmionic subsystems even allows for the detection of a topological contribution: While the outer ring deflects electrons into one transverse direction, the inner ring redirects electrons in the other direction. Since the detecting leads are of finite size one observes an oscillating Hall signal when the skyrmionium moves through them allowing for a highly reliable reading process.

Compared to skyrmions the main advantages of utilizing skyrmioniums as bits of information are (a) the slightly higher velocity (effect increases for narrower racetracks), (b) the absence of an acceleration phase ($v_x$ is instantly proportional to $j_x$; this effect is more prominent for a wider track), and (c) the skyrmionium moves always in the middle of the track. Advantages (b) and (c) are essential for an effective reading process, allow for changes in the moving direction and --- combined with the ultra-fast writing speed of the presented optoelectrical approach --- allow for a convenient `writing-while-moving' as well as `deleting-while-moving' functionality of a skyrmionium racetrack (cf. Fig.~\ref{fig:move_sksk}c).

In conclusion, writing and reading of magnetic skyrmioniums in thin films can be exploited to allow for the operation of an efficient skyrmionium-based racetrack storage device. In contrast to other magnetic quasiparticles, that are predicted to move without a skyrmion Hall effect, skyrmioniums have already been detected in experiments. Our proposals will expedite the development of a working data storage device based on magnetic quasiparticles.\\
\\
\textbf{Methods}\\
\textbf{Micromagnetic simulations.}
We use the GPU-accelerated micromagnetic software package Mumax3~\cite{vansteenkiste2011mumax,vansteenkiste2014design} to solve the LLG equation with the SOT term for every magnetic moment $\vec{m}_i$ of the discretized magnetization~\cite{landau1935theory,gilbert1955lagrangian, slonczewski1996current}
\begin{align}
\dot{\vec{m}}_i=&-\gamma_e\vec{m}_i\times\vec{B}_{i,\mathrm{eff}}+\alpha\vec{m}_i\times\dot{\vec{m}_i}\\
&+\gamma_e \epsilon\beta[(\vec{m}_i\times\vec{s})\times\vec{m}_i].\notag
\end{align}
Here, $\gamma_e=1.760\times 10^{11}\mathrm{T}^{-1}\mathrm{s}^{-1}$ is the gyromagnetic ratio of an electron. The in-plane torque coefficient is $\epsilon\beta=\frac{\hbar j\Theta_\mathrm{SH}}{2ed_zM_s}$; the out-of-plane torque parameter is set zero as it is small and does not drive the quasiparticles. The space- and time-dependent effective magnetic field 
\begin{align}
\vec{B}_{\mathrm{eff}}^i=-\frac{\delta F[\vec{m}]}{M_s\delta \vec{m}_i} \label{eq:beff}
\end{align}
is derived from the system's total free energy density $F$, given as the sum of exchange interaction, magnetocrystaline anisotropy, the demagnetization field, Zeeman energy, and DMI.

To generate skyrmioniums we use a modified photosensitive switch setup as shown in Fig.~\ref{fig:skyrmionium}a, motivated by the experimental results from Yang \textit{et al.}~\cite{yang2017ultrafast}. As discussed in the main text an optoelectrically induced spin current is superposed on the uniform spin current, which drives the skyrmioniums along the racetrack. For the simulations we assumed a Gaussian envelope in time according to Ref.~\onlinecite{yang2017ultrafast}.

Additionally, for the room temperature simulations an effective thermal field is included as 
\begin{align}
\vec{B}_\mathrm{therm}^i=\vec{\eta}\sqrt{\frac{2\alpha \mu_0 k_\mathrm{B}T}{M_s\gamma\Delta V \Delta t}}\, ,
\end{align}
where $\vec{\eta}$ is a random vector generated according to a standard normal distribution for each simulation cell and changed after every time step. $k_\mathrm{B}$ is Boltzmann's constant, $T$ the temperature, $\Delta V$ the simulation cells' size and $\Delta t$ the time simulation's step. 
The thermal fluctuations due to the room-temperature ambience lead to deformations of the skyrmionium structure, but the switching mechanism still works successfully. 

The system of Co/Pt is described by the following parameters~\cite{zhang2016control,sampaio2013nucleation}: saturation magnetization $M_\text{s}=0.58\,\mathrm{MA}/\mathrm{m}$, exchange stiffness $A=15\,\mathrm{pJ}/\mathrm{m}$, interfacial DMI $D=3.5\,\mathrm{mJ}/\mathrm{m}^2$, uniaxial anisotropy in $z$-direction $K_z=0.8\,\mathrm{MJ}/\mathrm{m}^3$, Gilbert damping parameter $\alpha=0.3$ and the spin Hall angle $\Theta_\mathrm{SH}=0.4$. We simulate a Co nanowire racetrack of width $150\,\mathrm{nm}$ and thickness $d_z=1\mathrm{nm}$, and discretize the magnetization in cubic cells of size $1\,\mathrm{nm}^3$.

\bcor We use these values for comparability with Refs.~\cite{zhang2016control,sampaio2013nucleation}, while noting that the DMI constant \cite{simon2018magnetism} and the spin Hall angle \cite{tao2018self} are still under debate. For the here presented parameters a skyrmionium is stable for DMI strengths between $3.3\,\mathrm{mJ}/\mathrm{m}^2$ and $3.7\,\mathrm{mJ}/\mathrm{m}^2$ (cf. Supplementary Figure~S5). In this context we note, that the effective DMI constant can be tuned, for instance as in a Pt/Co/Ir setup presented in Ref.~\cite{moreau2016additive}  or by utilizing a different bilayer system, what is possible since our predictions are generally applicable and not limited to Co/Pt interfaces.\ecor\\
\\
\textbf{Topological Hall effect calculations.}
To calculate the topological Hall resistivity we consider the tight-binding Hamiltonian (Eq.~\ref{eq:ham}) on a finite square lattice that forms the racetrack, as in Fig.~\ref{fig:read}. We apply four leads to the track: to the left and right to inject a small reading current, i.\,e., $I_l=-I_r=I$ and $V_l=-V_r$, and up and down to detect the voltage due to the transverse deflection and accumulation of the electrons, i.\,e., $I_u=I_d=0$ and $V_u$ and $V_d$. The transverse resistance follows directly from these voltages and currents, see text. To calculate the relationship between the currents and voltages we use a Landauer-B{\"u}ttiker approach~\cite{landauer1957spatial,buttiker1988absence}, by analogy with Refs.~\onlinecite{hamamoto2016purely,yin2015topological}, where skyrmions have been investigated. For the calculations we use the transport simulation package Kwant~\cite{groth2014kwant}.

We solve the set of linear equations $\{m,n\}=\{l,r,u,d\}$
\begin{align}
I_m=\frac{e^2}{h}\sum_n T_{mn}V_n,
\end{align} 
containing the transition matrix 
\begin{align}
T_{mn}=\mathrm{Tr}(\Gamma_m G_{mn} \Gamma_n G^{\dagger}_{mn}),
\end{align}
for the current $I$ and non-fixed voltages $V_u$ and $V_d$.
Here, the retarded Green's function 
\begin{align}
\vec{G}=\left(E-H-\sum\Sigma_i\right)^{-1}
\end{align}
and $\Gamma_i=\mathrm{i}(\Sigma_i-\Sigma_i^\dagger)$ enter ($E$ energy, $H$ tight-binding Hamiltonian). $\Sigma_i$ is the self energy of the $i$th lead.

Analyzing the results for different geometric parameters we find that skyrmioniums need to have a minimal size so that the topological charge density is well resembled. The leads should not be too large (optimally below half the skyrmionium radius) since they integrate the locally distinct signal making it broader and ambiguous. The distance between two bits can be small but then their signals begin to overlap, hampering an unambiguous detection. A minimal distance is given by $2(r_0+x_0)$, which is the width of the predicted signal (orange). In `0' bit regions oscillations of the signal around zero are visible originating from backscattering of electrons from the racetrack edges. This unfavorable effect decreases for wider tracks. \\
\\
\textbf{Acknowledgements}\\
This work is supported by Priority Program SPP 1666, CRC/TRR 227 and SFB 762 of Deutsche Forschungsgemeinschaft (DFG).\\
\\
\textbf{Author contributions}\\
B.G. and A.S initiated research and planned the project. J.B, I.M and S.P. supervised the project. B.G. conducted calculations of the topological Hall effect and analyzed the Thiele equation. A.S. did the micromagnetic simulations. B.G. and A.S. wrote the manuscript. All authors discussed the results and commented on the manuscript.\\
\\
\textbf{Competing interests}\\
The authors declare no competing interests.



\end{document}